\documentclass{article}

\usepackage{color}
\usepackage{times}
\usepackage{epsfig}
\usepackage{graphicx}
\usepackage{amssymb}
\usepackage{amsmath}
\usepackage{amssymb}
\usepackage{hyperref}
\usepackage{stfloats}
\usepackage{placeins}
\usepackage{caption}
\usepackage{multicol}
\usepackage[margin=1in]{geometry}

\definecolor{willcolor}{RGB}{23, 127, 117}
\definecolor{evacolor}{RGB}{23, 127, 117}
\definecolor{bcolor}{RGB}{182, 33, 45}

\title{From sample to knowledge:  Towards an integrated approach for neuroscience discovery}

\author{William Gray Roncal$^{1,2,*}$, Eva L Dyer$^{3,4,*}$, Doga G\"ursoy$^{5}$, Konrad Kording $^{3,4}$,\\ Narayanan Kasthuri $^{6,7}$ 
\\ \\
$^{1}$Johns Hopkins University, Department of Computer Science, Baltimore, MD\\
$^{2}$JHU Applied Physics Laboratory, Research and Exploratory Development, Laurel, MD \\
$^{3}$ Dept. of Physical Medicine and Rehabilitation, Northwestern University, Chicago, IL \\
$^{4}$ Sensory Motor Performance Program, Rehabilitation Institute of Chicago, Chicago, IL \\
$^{5}$ Advanced Photon Source, Argonne National Laboratory, Lemont, IL \\
$^{6}$ Center for Nanoscale Materials, Argonne National Laboratory, Lemont, IL \\
$^{7}$ Dept. of Neurobiology, University of Chicago, Chicago, IL \\ \\
$^{*}$ Authors contributed equally. \\
Corresponding Authors:  William Gray Roncal and Eva L. Dyer, \\email:\{wgr@jhu.edu, edyer@northwestern.edu\} \\
}
\begin{document}

\maketitle

\begin{abstract} 

Imaging methods used in modern neuroscience experiments are quickly producing large amounts of data capable of providing increasing amounts of knowledge about neuroanatomy and function. A great deal of information in these datasets is relatively unexplored and untapped. One of the bottlenecks in knowledge extraction is that often there is no feedback loop between the knowledge produced (e.g., graph, density estimate, or other statistic) and the earlier stages of the pipeline (e.g., acquisition). We thus advocate for the development of sample-to-knowledge discovery pipelines that one can use to optimize acquisition and processing steps with a particular end goal (i.e., piece of knowledge) in mind. We therefore propose that optimization takes place not just within each processing stage but also between adjacent (and non-adjacent) steps of the pipeline.  Furthermore, we explore the existing categories of knowledge representation and models to motivate the types of experiments and analysis needed to achieve the ultimate goal. To illustrate this approach, we provide an experimental paradigm to answer questions about large-scale synaptic distributions through a multimodal approach combining X-ray microtomography and electron microscopy.

\end{abstract}

\section{Introduction}
\label{sec:intro}

Much of scientific exploration involves three main stages to translate raw data to knowledge suitable for making scientific discoveries:  acquisition, processing, and analysis.  These stages, which begin with sample collection and result in mathematical analysis (knowledge), are typically performed independently. Thus, when scientists optimize any of these stages, the process is typically done in a feed-forward manner, where at each stage one does not have the ability to revise previous steps.  This is problematic because it is important to consider the best set of parameters from a global context---the question of interest might well lead to a solution that is not obvious at a particular stage of the pipeline. Many challenges and potential improvements have been identified (e.g., \cite{Plaza2014,Lichtman2014}) - by combining these ideas with an integrated approach, we believe that significant advancements may be made.

For questions in the sub-discipline of brain imaging science, the stages outlined above generally include obtaining and imaging the specimen (e.g., microscopy), processing the data to reveal and identify relevant structures (e.g., alignment, registration, computer vision, image processing), and finally extracting knowledge. In the first step, a specific preparation and imaging technique are used to produce measurements (image data) from the sample. We note that image data is typically calibrated to produce the best image quality, as judged visually or by a surrogate metric. Second, these images are processed to estimate or detect the relevant neural structures of interest. From this result, we build an abstraction or model of the data (e.g., graphs, density estimate, point process model). Finally, these models and abstractions that represent knowledge are used for analysis leading to scientific discovery.
 
\begin{figure}[h!]
\centering{
\includegraphics[width=1\textwidth]{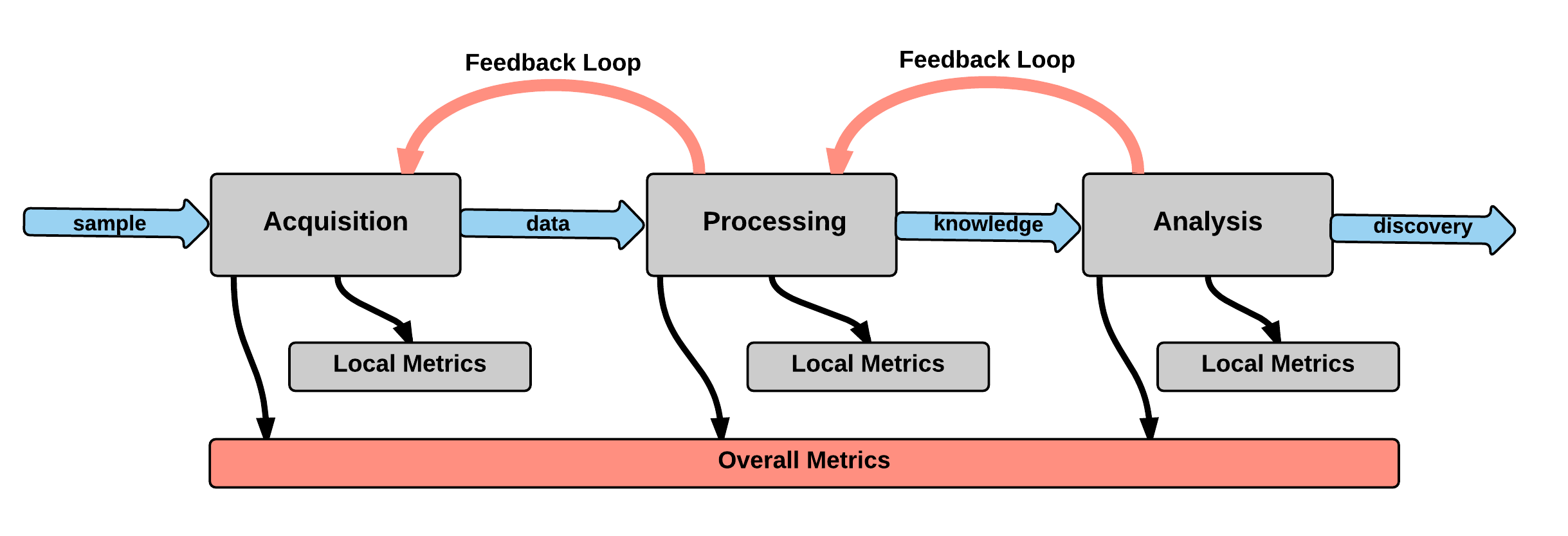}
\caption{ \small {Anatomy of an experiment:} \em  We illustrate our augmented processing sample-to-knowledge workflow for a scientific experiment.  A traditional workflow consists of a feed-forward chain of stages (gray), which represent major (often disparate) building blocks.  The products of these stages (blue arrows) represent the current interface points.  Our augmented pipeline adds feedback loops between stages and interfaces to an overall knowledge metric which may lead to improved performance. \label{fig:fig1}}
}
\end{figure}

As scientific questions are identified and explored, we believe that it is critical to design end-to-end experimental pipelines that are optimized with respect to metrics that are tailored to the ``knowledge objective'' we are interested in.  To this end, we must  develop metrics that provide us information about how close we are to the best approximation of our data with respect to a given model. Surrogate metrics are often used - these local measures of quality control may be useful for optimization, but may not map smoothly to the overall objective of interest.  Therefore, it is important to carefully choose these metrics and assess their impact on overall performance. Finally, in optimizing these metrics, we advocate for feedback loops at each stage of this pipeline and for a global optimization process.  In Figure 1, we outline this general approach of sample to knowledge, including modular and overall knowledge metrics.  In our setting, feedback loops exist to allow better imaging methods to be developed in conjunction with algorithm development; between analysis methods and processing methods; and by chaining these feedback loops to connect analysis and imaging.

\section{Defining Knowledge in Neuroanatomy}

As researchers produce estimates of brain connectivity that span multiple modalities, increasingly larger volumes, and multiple subjects, we need efficient abstractions and computational models to summarize the structure in these data. With sufficient descriptions and models of neuroanatomical samples (e.g., cells and their processes, connectivity measures, and blood vessels), neurocartographers can then apply  analysis techniques to effectively answer scientific questions leading to fundamental discovery in processing mechanisms and the structure of disease. 

Here, we will argue that each abstraction (or model) that we must estimate, whether it be the edges in a graph or an estimate of the density of synapses, introduces natural metrics for which we can measure the quality of an estimator.  When selecting a metric, we seek one that has a smooth mapping between the sample and knowledge produced; that is, small perturbations in data should produce small perturbations in the estimated quality of the knowledge produced. As an example, small variations in acquisition stage parameters might not produce a smooth mapping to downstream knowledge---potentially leading to deleterious impact on a new test sample.  We argue that given a particular question, imaging data and processing stages should be optimized with respect to the chosen metric on a per-problem basis. We envision end-to-end systems which can be adapted and optimized based upon the scientific inquiry of interest, and describe steps toward this vision.

While not by any means exhaustive, to make our exposition more concrete, we now describe two examples of knowledge that are of interest in neuroanatomy.

\subsection{Graphs}

Graphs are a mathematical abstraction (model) that has been heavily explored in the context of both functional and structural maps of cortex. A graph, defined as $G = (V, E, A)$ consists of vertices ($V$) (modeling discrete objects) and their edges ($E$) (connections), which may refer to different measures at different scales and contexts (e.g., synapses in EM, functional connections in fMRI).  Graphs can have attributes ($A$) on both vertices and edges, providing information such as connectivity weight and direction.

\subsubsection{Knowledge to extract from graphs} 
Once we have estimated a graph, there are a number of analyses that can be performed. Studying the ``community structure'' and motifs in a graph has been popularized in the past decade, and a number of works have studied clustering in neural connectivity graphs (e.g.,  \cite{Lyzinski2015}).


Understanding the impact that a change to the structure of a graph has on a particular function on the graph is still challenging. Indeed, many functions that we wish to compute on a graph, such as the clustering coefficient and the degree distribution, do not produce smooth mappings when we perturb our estimate of the graph structure. Thus, when we change our estimate of a graph, we can produce dramatic changes in the sufficient statistic of interest.  Therefore it is especially critical to consider the inference impacts on graph changes when designing robust and efficient metrics.
When considering errors in connectomes, several ideas have been explored, including treating connections between synapses (i.e., the line-graph) as a detection problem and optimizing the f1-score \cite{GrayRoncal2015a}, as well as quantity-quality tradeoffs \cite{Priebe2012}.

\subsection{Spatial point processes} 

As many objects of interest in the brain, such as cells and synapses, are distributed spatially and detected in images as discrete objects, one can model a collection of such objects as a spatial point process, either in 2D or 3D \cite{baddeley2007spatial,prodanov2007spatial,hansson2013ripleygui}. A point process can be used to describe the occurrence of an object (cell) at a particular spatial position (x,y,z). Thus the spatial location of an object is an important aspect of the model.


We consider a process marked when we have additional information associated with each occurrence. For instance, in our recent study we compute the radius, confidence estimate, local density of each detected cell \cite{xbrain2016}. All of these features are considered marks, which summarize information about each detected object.  

\subsubsection{Knowledge to extract from point process models}

{\em Density estimation.} While histogram estimates are a traditional tool used to obtain a measure of density, as the dimensionality grows, histograms become exceedingly poor indicators of density. Even in low dimensions, when we oversample the bin size for a histogram, the histogram varies wildly, even when the distribution varies smoothly. In addition, histograms are known to oversmooth the data when the bin size is too large.

Instead of histogram approaches, nonparametric methods that are based upon nearest neighbor relationships between detected objects (cells) can be used. Density estimation is an active area of research in machine learning and statistics, where kernel density estimators are considered the state-of-the art.  Using density estimates, we can compute the correlation between density at a given spatial location and the size or distance of the nearest vessel.  We can also assess variance in density as a function of a local region in volume (or layer).  A final example is to consider the degree to which counts are Poisson (Fano factor of spatially varying 3D point process).  To optimize this knowledge representation, KL-divergence between density estimates or precision/recall of the detection or segmentation results may be used.  These ideas have been explored in various contexts in neuroscience:  Cell and vessel densities in light microscopy (optical sectioning) and $\mu$CT \cite{xbrain2016}, as well as synaptic densities in electron microscopy \cite{Anton-Sanchez2014}.  






\section{Sample to Knowledge Pipelines}

To our knowledge there are few examples of fully integrated sample-to-knowledge pipelines in neuroimaging, although much research has focused on building well-engineered stages that optimize a chosen local metric.  We assert that these stages can be readily combined to produce a true sample-to-knowledge pipeline, leading to improved efficiencies and performance.  These processing modes seem to be a common motif - existing across many groups and modalities.


\subsection{Imaging Pipelines}
Neuroimaging researchers have developed many different pipelines to produce image reconstructions.  In electron microscopy data \cite{Kasthuri2015,Bock2011}, these pipelines often produce large image tiles that can be aligned to produce a volume \cite{Bria2012,Saalfeld2010}.  In x-ray $\mu$CT, we are able to directly reconstruct the volume after selecting a variety of parameters \cite{xbrain2016}. In MRI-based connectomics, brains are imaged in-vivo using a strong magnetic field to produce structural and diffusion weighted image volumes\cite{Bihan1986}.  Here we consider the output of the imaging pipeline to be an aligned volume.  Imaging researchers may optimize this stage without explicitly considering potential downstream processing (e.g., manual labeling approaches versus automated segmentation algorithms).  Making this connection and feedback loop part of the scientific process allows for data and algorithms to be developed jointly.

\subsection{Processing Pipelines}
Machine learning and computer vision researchers have developed many processing algorithms and pipelines to transform image volumes to a knowledge representation.  For example, in EM connectomics, typically cell membrane pixels are found; this information is used to create supervoxels; and finally the supervoxels are agglomerated using various affinity measures \cite{Plaza2016}.  This information is combined with (separately) detected synaptic connections to form a graph.  In our work with x-ray tomography, we segment blood vessels and detect individual cell bodies to obtain density estimates.  In structural MR connectomics, typically labeled regions from MPRAGE data are combined with fiber tracts inferred from diffusion weighted images to estimate a graph (e.g., \cite{Daducci2012}). Many other pipelines exist for processing neuroscience data (e.g., \cite{Kaynig2013b,Plaza2016}), including several that span processing and analysis (although not acquistion)\cite{GrayRoncal2015a,Freeman2014}.  Computer vision researchers may be tempted to treat the unique challenges of neuroimaging as a black-box challenge and deal with only feature vectors and labels.  Doing so risks ignoring important global contextual queues.  Classifier robustness and generalization to diverse data may be greatly enhanced with a more integrated approach.

\subsection{Analysis Pipelines}

The analysis pipeline will vary depending on the question of interest and often proceeds according to standard statistical methodologies.  When this analysis is performed in isolation, researchers may lack the domain knowledge to avoid data irregularities.  Furthermore, better processing decisions could be made with increased awareness of underlying statistical assumptions.  A common example of processing bias is constructing a graph based on a computer-vision chosen operating point.  For example, the processing researcher might arbitrarily choose to optimize the harmonic mean of precision and recall (the f1-score).  This choice is likely to be suboptimal if chosen without knowlege of the inference model or statistical assumptions.

\subsection{Hyperparameters}
Each step in the overall sample-to-knowledge pipeline requires selecting values for many different parameters.  Many are explicit (e.g., amount and type of post-staining during image acquisition, minimum number of keypoints in alignment, precision-recall operating point).  Others are implicit and relate to more fundamental experimental design choices.  Enumerating all of these options leads to a large space of hyperparameters which can be optimized both globally and locally according to a chosen metric.  

\subsection{Optimization}

Naively, one could jointly optimize each parameter through an exhaustive search.  However, this method is combinatorally hard and quickly becomes intractable as the number of parameters grows.  As an illustration, if each of 100 parameters had 5 possible values across the entire experimental setup, $5^{100} \approx 10^{70}$ options would have to be evaluated, where each evaluation requires running the entire experiment.

Some dimensionality reduction is clearly possible via subsystem optimization and intuition.  However, the remaining parameter space is likely to still be very large, and can be addressed by local optimization strategies or more sophisticated global optimization strategies.  Recently, some of the authors have developed a novel approach for hyperparameter optimization that scales well with dimension \cite{gheshlaghi2016convex}. Employing these (and other) methods for global optimization will enable us to find the best operating point for a pipeline across all stages of the processing chain.

Finally, it is important to design tools in a modular way, preferably in accordance with open-science, to allow for the comparison of different knowledge extraction protocols.  Our research group leverages tools from NeuroData \cite{Burns2013}.

\subsection{Automatic whole brain segmentations}
Within neuroscience, there are agreed upon ways of dividing the brain into structures and regions. For example, Brodman defined a way of dividing cortex into many regions. Such segmentations are important as they define a way of quantifying the effect of diseases and of function. We are working on utilizing information from large scale brain imaging to divide the brain into such regions using automatic algorithms. Combinations of spectral clustering, with Bayesian methods and fast sparse matrix methods are most promising.
\section{Case Study}

Here we propose an experimental paradigm (Figure 2) leveraging advances in  multi-modal brain mapping approaches (e.g., electron and x-ray microscopy).  Our goal is to better estimate synaptic density and investigate community structure through strategic sampling.  We will detail how large mesoscale maps of brains can complement higher resolution EM maps and may potentially bridge the scale gap.  
 
\subsection{Experimental Design}

Our basic approach is to (1) identify methods and steps required for each stage of the pipeline, (2) identify implicit and explicit parameters for each method or algorithm, (3) outline a plan for analysis and metrics.

\begin{figure}[h!]
\centering{
\includegraphics[width=0.95\textwidth]{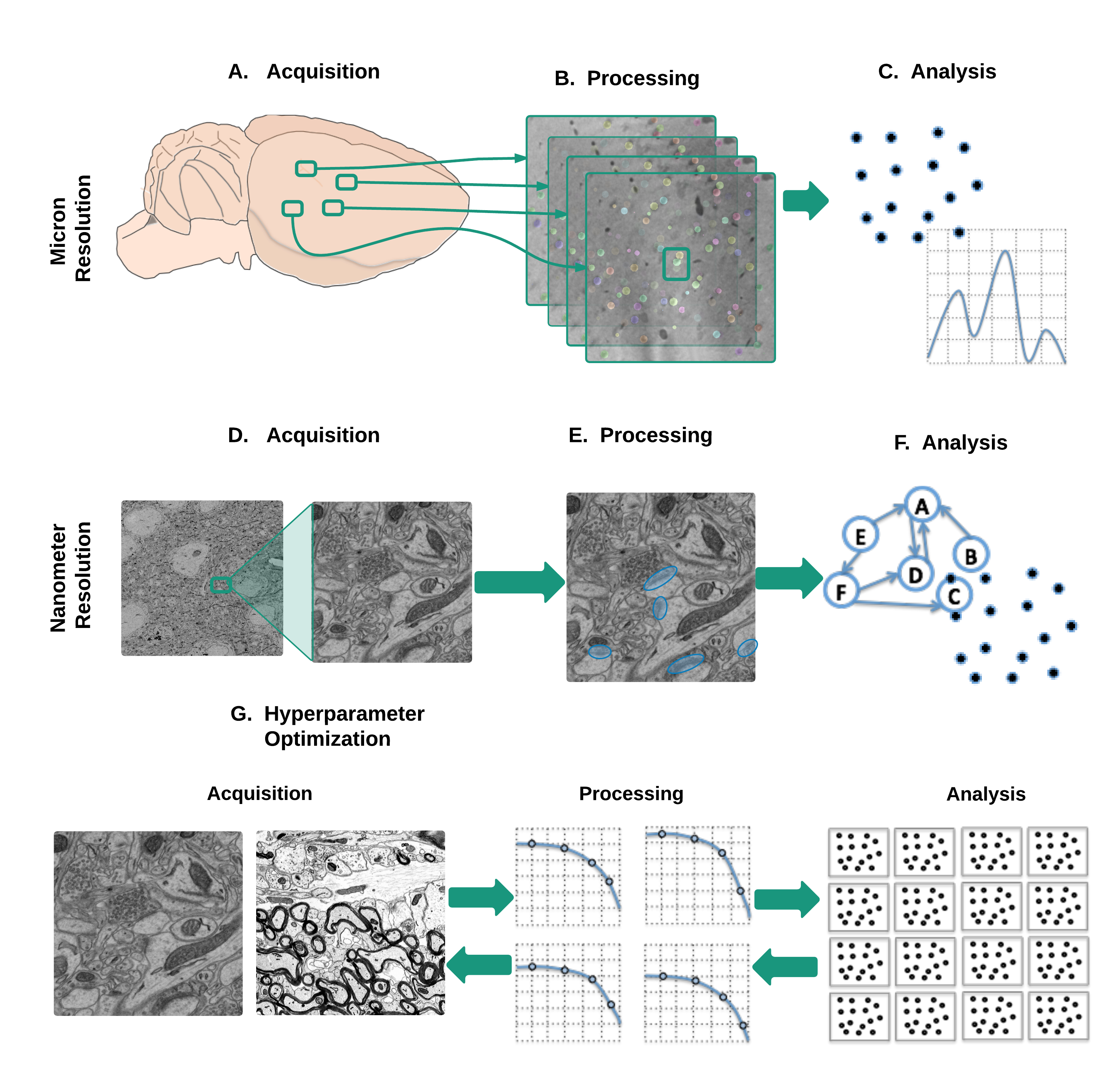}
\caption{ \small {Multimodal Synapse Motifs} \em   This figure illustrates the experimental paradigm proposed in this manuscript for a particular problem of scientific interest.  This experimental design actually contains two interlinked sample-to-knowledge pipelines; each panel illustrates the expected result from that stage.  (A) Initially a mesoscale image of the brain is reconstructed using X-ray microtomography.  (B) This image volume is then used by computer vision algorithms to find an estimate of cell body location and size.  (C) This knowledge can be represented as a map of the cell body locations in space, along with relevant attributes such as confidence and size.  (D) High resolution electron microscopy imaging occurs for blocks selected during the previous step.  X-ray imaging is non-destructive, and so it is possible to re-image interesting locations in the same sample.  (E) Next, we locate all synapses using automated approaches, which leads to (F) knowledge about relative position and densities for each block in support of scientific discovery.  (G) Finally, we observe that at each stage opportunities exist for local and global optimization of knowledge.  This illustrates both the exchange of information between stages and the rapid explosion of possible knowledge realizations.\label{fig:fig2}}
}
\end{figure}

\begin{itemize}
\item {\em Acquisition (X-ray microtomography):} Images are acquired and reconstructed using synchotron X-Ray micro-tomography and tomoPy image reconstruction software \cite{gursoy2014tomopy} as described in \cite{xbrain2016}.

\item {\em Processing (Cell-Detection (X-ray):}  Soma are identified using ilastik \cite{Sommer2011}, a random forest-based classification toolkit, combined with a greedy sphere finding algorithm.  Size estimates are refined by examining the probability support under each detection \cite{xbrain2016}.

\item {\em Analysis (X-ray):}
Based on these detections, local and global density (or other metrics) can be used to select interesting regions to image under a higher resolution electron microscope.

\item {\em Image Reconstruction (EM):}
In electron microscopy data, tissue is fixed, stained, sliced, and imaged following a protocol similar to \cite{Kasthuri2015}.  

\item {\em Processing (EM):}  Various techniques exist for synapse detection \cite{Becker2012a,Kreshuk2014,GrayRoncal2015b} and options will need to be evaluated depending on computational requirements and sample preparation.

\item {\em Analysis (EM):}
Given local density and structure estimates of synapse locations, it will now be possible to compute whole brain density estimates and look for local substructure.
\end{itemize}

\subsection{Hyperparameter Optimization}
\label{sec:hyperp}
This section briefly highlights some of the parameters chosen at each step to give a sense of the design space and complexity.  This list highlights some of the more impactful options, but is by no means exhaustive.  Because X-ray imaging is non-destructive, it is possible to image the same sample at both micron and nanometer resolution.  For many variations in parameters (e.g., X-ray reconstruction, precision-recall operating points), we can use the same sample and sometimes even the same training labels, aiding our search for optimal processing.

\begin{itemize}
\item {\em Acquisition (X-ray):} Examples of relevant hyperparameters include the choice of phase retrieval filter used in slice reconstruction, filter bandwidth, filter cutoff, ring removal parameters, median filtering, and contrast enhancement settings.

\item {\em Processing (Cell-Detection (X-ray)} Examples of relevant hyperparameters include the classifier type and parameters, number of classes, features and feature parameters, choice of cell-finding algorithm, estimated cell size, number of iterations, and precision-recall curve operating point.

\item {\em Analysis (X-ray)} Examples of relevant hyperparameters necessary to compute density estimates include the number of nearest neighbors, bin size, and the method used.

\item {\em Image Reconstruction (EM)"} Examples of relevant hyperparameters include the fixation protocol, type and quantity of stain, imaging resolution, alignment algorithm, alignment degrees of freedom, color correction, histogram normalization.
 
\item {{\em Processing (EM):} Examples of relevant hyperparameters include the choice of synapse detection algorithm, image features, the choice of training volume, labels used for training, operating point on precision recall curve.}

\item{{\em Analysis (EM):} Examples of relevant hyperparameters include the window-size, method to compute density estimate, sample point weighting, outlier rejection and masking, and the  statistical test used to compare regions.}
\end{itemize}

\subsection{Analysis and Metrics}

Choosing correct metrics is an open question, although we highlight three specific ideas for this particular problem.  First, we propose to explicitly evaluate the link between image reconstructions and cell or synapse detection by  choosing image parameters through qualitative or local metrics and creating a series of detection precision-recall curves.  Second, we plan to choose different points on Pareto front of the family of precision-recall curves and examine the impact on the model in terms of quality and smoothness.  Finally, we will compare the overall analysis estimates derived from our knowledge to known scientific discoveries for context and validation.

\subsection{Discussion}

Once we have generated knowledge, we have all of the raw inputs needed for discovery.  However, it is still a research question to find the best way to model knowledge to promote discovery.  Additionally, finding metrics that most accurately model errors in our knowledge estimates will lead to better inference and a lower-dimensional hyperparameter space.  

Multi-modal neuroimaging studies hold great promise for information fusion and rapidly building brain maps at scales that seemed impossible until very recently.   It is now possible to create structural brain maps across at least six orders of magnitude, ranging from millimeter pixel resolution in magnetic resonance imaging to nanometer pixel resolution in electron microscopy.  These methods each offer different information and complexity and may lead to combined multimodal, multiresolution maps.  However, the additional complexity requires thoughtful sampling, analysis, and sample preparation to maximize information gain.  Although this manuscript focuses on structural connectomics, many of the principles apply to functional maps like calcium imaging and fMRI and can be explored further.

In order to create knowledge from diverse datasets, we argue that it is essential to improve evaluation and metrics used across disparate research techniques and groups.  This will allow the community to develop the best pipelines and tools to efficiently extract and combine information.  Future work will explore the specific experiment outlined above to estimate synaptic structure at large scale.

\subsection{Acknowledgments}

This research used resources of the U.S. Department of Energy (DOE) Office of Science User Facilities operated for the DOE Office of Science by Argonne National Laboratory under Contract No. DE-AC02-06CH11357.  Partial support was received from a JHU Applied Physics Laboratory Educational Fellowship.

\bibliographystyle{IEEEtran}
\bibliography{perspectives.bib}	

\end{document}